\documentclass[prl,twocolumn,showpacs,preprintnumbers,amsmath,amssymb]{revtex4}
\usepackage{graphicx}
\usepackage{dcolumn}
\usepackage{bm}

\begin{document}
\title{Spin triplet superconducting state due to broken inversion symmetry in Li$_2$Pt$_3$B}
\author{M. Nishiyama $^{1}$,  Y. Inada $^2$ and Guo-qing Zheng $^{1}$}
\address{ $^1$ Department of Physics, Okayama University, Okayama 700-8530, Japan}
\address {$^2$ Faculty of Education, Okayama University, Okayama 700-8530, Japan }


\begin{abstract}
We report $^{11}$B and $^{195}$Pt NMR measurements in non-centrosymmetric superconductor Li$_2$Pt$_3$B. We find that the spin susceptibility measured by the Knight shift remains unchanged across the superconducting transition temperature $T_c$. 
With decreasing temperature ($T$) below $T_c$,  the spin-lattice relaxation rate $1/T_1$ decreases with no coherence peak and is in proportion to $T^3$. These results indicate that the Cooper pair is in the spin-triplet state and that there exist line nodes in the superconducting gap function. They are in sharp contrast with those in the isostructural Li$_2$Pd$_3$B which is a spin-singlet, $s$-wave superconductor, and are ascribed to the enhanced spin-orbit coupling due to the lack of spatial inversion symmetry. Our finding points to a new paradigm where exotic superconductivity arises in the absence of strong electron-electron correlations. 
\end{abstract}

 \hspace{5cm} Phys. Rev. Lett. {\bf 98}, 047002 (2007)

\maketitle


In most superconducting materials, there is an inversion center in the crystal which guarantees the parity conservation.
 In conventional superconductors, such as Al, where the Cooper pair is formed by the attractive force produced by lattice vibration, the orbital wave function (OWF) of the Cooper pairs is in the $s$-wave form. Since an electron must obey the Fermi statistics, the two spins of such Cooper pair must be in the singlet state. This is also true in most strongly correlated electron systems such as high transition-temperature ($T_c$) copper-oxides \cite{Tsuei}, cobalt oxide  Na$_x$CoO$_2$$\cdot$1.3H$_2$O \cite{Co} and many heavy-fermion compounds \cite{HF}, where the OWF is also symmetric  although it has nodes (zeroes). In contrast, if the OWF is asymmetric about the origin with nodes, {\it e.g.}, a $p$-wave function, the Cooper pair must be in the spin-triplet state. Such pairing state is realized in superfluid $^3$He  \cite{Lee} and also believed to occur in strongly correlated electron superconductors UPt$_3$  \cite{Tou}, Sr$_2$RuO$_4$  \cite{Ishida} and (TMTSF)$_2$PF$_6$  \cite{Brown}. 
 
 However, when a superconductor lacks a crystal inversion center, the above-described rule (parity conservation) is violated due to the asymmetric spin-orbit coupling (SOC), and the pairing symmetry becomes non-trivial \cite{Gorkov,Frigeri,Sa,Fr}. 

In this Letter, we present NMR (nuclear magnetic resonance) evidence  that increasing the strength of the SOC drastically changes the electron pairing symmetry in non-centrosymmetric superconductors Li$_2$Pt(Pd)$_3$B. The perovskite-like cubic compounds
Li$_2$Pd$_3$B and Li$_2$Pt$_3$B are superconducting at $T_c\sim$  7 K and $\sim$2.7 K, respectively \cite{Togano1,Togano2}.
 The inversion symmetry breaking effect is much larger compared to known compounds such as CePt$_3$Si (Ref.\cite{Bauer}); all the elements, including the heavy element  Pt(Pd), are located in non-centrosymmetric positions, while in CePt$_3$Si the main effect comes from non-centrosymmetric Si which is a much lighter element. Also, in  CePt$_3$Si or UIr (Ref. \cite{Akazawa}), the correlated $f$-electrons play a major role in determining the superconducting properties \cite{Yogi,Fujimoto}; note that the $4f^0$ analog of the former compound, LaPt$_3$Si, is a conventional superconductor \cite{Onuki}. However,  there are no electron correlations in Li$_2$Pd$_3$B \cite{Nishiyama,Yokoya}, which turns out to be also true in Li$_2$Pt$_3$B (see below).
 Li$_2$Pd$_3$B is  a spin singlet,   $s$-wave superconductor as we reported previously \cite{Nishiyama}. In Li$_2$Pt$_3$B where the SOC is much stronger, Yuan {\it et al} suggested very recently that their magnetic penetration depth data can be interpreted by  an extended $s$-wave, spin-triplet model  \cite{Yuan}. Here we present direct evidence from the measurement of spin susceptibility that the Cooper pair is in the spin triplet state in Li$_2$Pt$_3$B.  We also find that there exist line nodes in the OWF. These findings point towards a new paradigm where exotic superconductivity arises without electron-electron correlations. 

Poly-crystal samples of Li$_2$Pt$_3$B were prepared by the arc-melting method with starting materials of Li (99.9\% purity),  Pt (99.9\%) and B (99.5\%). The two-step arc melting process \cite{Togano1} was used. 
For NMR measurements, the sample was crushed into powder. 
$T_c$ at zero and a finite magnetic field ($H$) was determined by measuring the ac susceptibility using the in-situ NMR coil. $T_c(H=0)$ is 2.68 K. $H_{c2}$ was estimated to be  $\sim$1.5 T,  which is in  agreement with the report by Badica {\it et al}  \cite{Togano2}.
In order to minimize the reduction of $T_c$ by $H$, the measurements were done at very low fields of 0.26 T for $^{11}$B and 0.39 T for $^{195}$Pt. The NMR spectra were obtained by fast Fourier transform (FFT) of the spin echo taken at a constant $H$. The nuclear spin-lattice relaxation rate, $1/T_1$, was measured by using a single saturation pulse and by fitting the nuclear magnetization to a single exponential function since the quadrupole interaction is absent; the fitting is excellent. 
 Measurements below 1.4 K were carried out in a $^3$He-$^4$He dilution refrigerator. Efforts were made to avoid possible heating by the RF pulse,  such as using a small-amplitude and low-frequency (low $H$)  RF pulse.

The most direct probe for spin pairing state  is the spin susceptibility $\chi_s$ via the measurement of  the Knight shift. If the Cooper pair is in the singlet state, $\chi_s$ will vanish at $T << T_c$. For triplet pairing, however, $\chi_s$  will remain unchanged across $T_c$. Figure 1 shows the NMR spectra, with those of   Li$_2$Pd$_3$B for comparison.
Figure 2 shows the low-$T$ blow-up of the measured $^{11}$B Knight shift. The observed Knight shift ($K_{obs}$) is composed of the spin part ($K_s$) and the orbital part ($K_{orb}$).    $K_{orb}$ is $T$ independent, and  $K_s$ is proportional to  $\chi_s$,   
$K_{s} = A_{hf}\chi_s$,            
where $A_{hf}$ is the hyperfine coupling between the nuclear and electron spins. We first recall the data for Li$_2$Pd$_3$B where the shift increases  below $T_c$ ($H$=1.46 T) =5.7 K  \cite{Nishiyama}, as can be seen in Fig. 2. This indicates the decrease of $\chi_s$  in the superconducting state, since $A_{hf}$ due to $p$-electrons of boron is negative \cite{Nishiyama}. Thus the spin pairing in Li$_2$Pd$_3$B is in the singlet state. The solid curve in Fig. 2 is a fit to the BCS theory,  which yields $K_{orb}$=0.085\% for $^{11}$B.

\begin{figure}
\begin{center}
\includegraphics[scale=0.50]{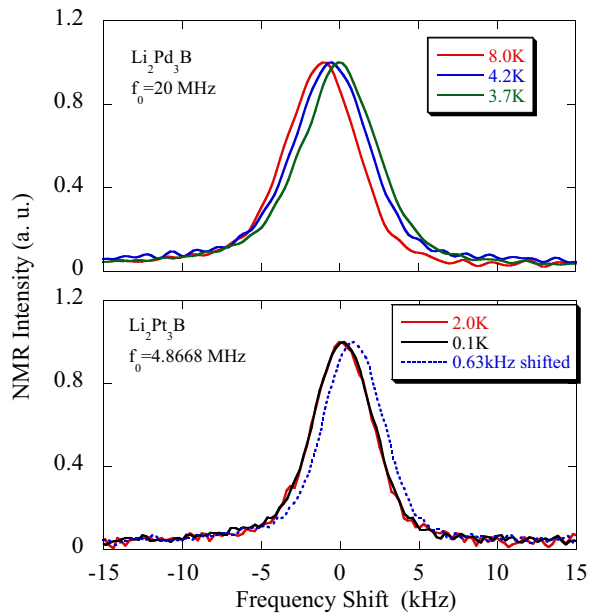}
\caption{(Color online) The temperature evolution of the $^{11}$B NMR spectra. The horizontal axis (shift) is measured with respect to Larmor frequency ($f_0$). In the lower panel for Li$_2$Pt$_3$B, the dotted curve depicts the expected spectrum at $T$=100 mK if the Cooper pairing is in the singlet form (see text).}
\label{fig:1}
\end{center}
\end{figure}

In contrast, the shift for Li$_2$Pt$_3$B does not change across $T_c$($H$=0.26 T)=2.1 K or $T_c$ ($H$ =0.35 T)=1.8 K. The dotted curve illustrates the behavior at $H$=0.26 T if the Cooper pair is in the spin-singlet state as in Li$_2$Pd$_3$B. As can be seen in Fig. 1 which shows the detailed evolution of the $^{11}$B NMR spectra with temperature,  the contrasting behavior in the two materials is evident.  In Li$_2$Pd$_3$B, the spectrum moves toward high frequency progressively below $T_c$. However, in Li$_2$Pt$_3$B, the spectra at $T$=2.0 K (above $T_c$) and 0.1 K (below $T_c$) are almost overlapping with each other. The dotted curve depicts the expected spectrum at $T$=0.1 K assuming spin-singlet pairing (namely, $K_s$=-0.013 \% vanishes completely), which should be shifted by 0.63 kHz. It is obvious that such shift change is much larger than the experimental resolution as can be judged from the upper panel of Fig. 1 for Li$_2$Pd$_3$B, and should be detectable if it would occur. 

These results indicate that the Cooper pair is in  a {\it spin-triplet} state in Li$_2$Pt$_3$B, since extrinsic cause for the contrasting behavior of the Knight shift can be excluded as elaborated below. First, one may worry about the influence of the vortices. However, $H/H_{c2}$ is 0.17 in Li$_2$Pt$_3$B, which is much smaller than 0.32 in Li$_2$Pd$_3$B \cite{Nishiyama,Togano2}. Therefore, the contribution from the normal electrons in the vortex cores, if any, is much smaller in Li$_2$Pt$_3$B.  Also, note that the shift at  $H$=0.26 and 0.35 T, and also 0.39 T (see below) shows  the same behavior. Second, one cannot ascribe the invariance of $K_s$ to the spin-orbit scattering due to impurities. 
 The full width at half maximum of the NMR spectra is about 3 gauss, which can be accounted for by dipole-dipole interaction alone, thus indicating that the sample is very clean.  Estimate from the residual resistivity of 25$\mu\Omega$$\cdot$cm in our polycrystal reveals a conservative mean free path $l_{tr}$$\sim$500 $\AA$. On the other hand, if the impurity scattering is so strong as to give, say, 90\% of the normal-state shift $K^N$, then the spin orbit mean free path $l_{SO}$ would be $\sim$23 $\AA$,  as can be estimated from the  formula $K^S/K^N=1-2l_{SO}/\pi\xi_0$ \cite{Anderson}, by using the coherence length  $\xi_0$=145 $\AA$. However, this is clearly inconsistent with the criterion of $l_{SO}$ $>>$ $l_{tr}$ requied for the above formula.
 Also, strong random scattering would  destroy the nodes in the gap function, which is imcompatible with our finding of clear nodes by $T_1$ measurement as described below.

\begin{figure}
\begin{center}
\includegraphics[scale=0.5]{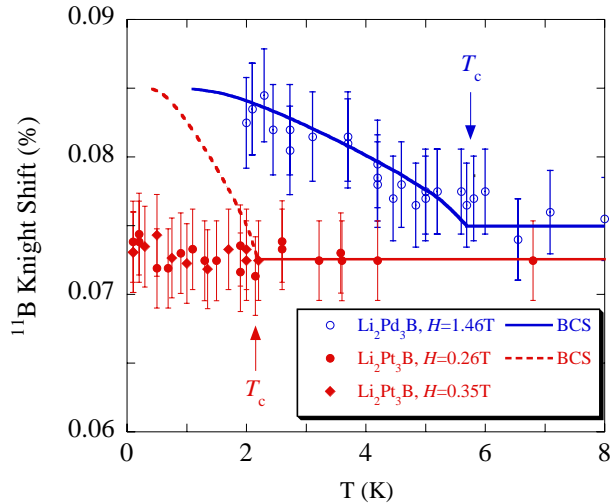}
\caption{(Color online) The $^{11}$B Knight shift at low temperatures for Li$_2$Pd$_3$B and Li$_2$Pt$_3$B. The curves are calculated $T$-dependence for spin-singlet pairing using the energy gap $\Delta_0=1.1k_BT_c$. $K_{orb}$ due to the orbital susceptibility is estimated to be 0.085\% (also see the main text). The arrows indicate $T_c$ ($H$=1.46 T)  for Li$_2$Pd$_3$B and $T_c$ ($H$=0.26 T) for Li$_2$Pt$_3$B, respectively.}
\label{fig:2}
\end{center}
\end{figure}

The insight into the Cooper pair's OWF can be gained from the $1/T_1$. As is reproduced  in Fig. 3, $1/T_1$ in Li$_2$Pd$_3$B is enhanced just below $T_c$ over its normal-state value, forming a so-called coherence peak \cite{Nishiyama}, which  is a hallmark of an isotropic superconducting gap \cite{Slichter}.
 
 

\begin{figure}
\begin{center}
\includegraphics[scale=0.5]{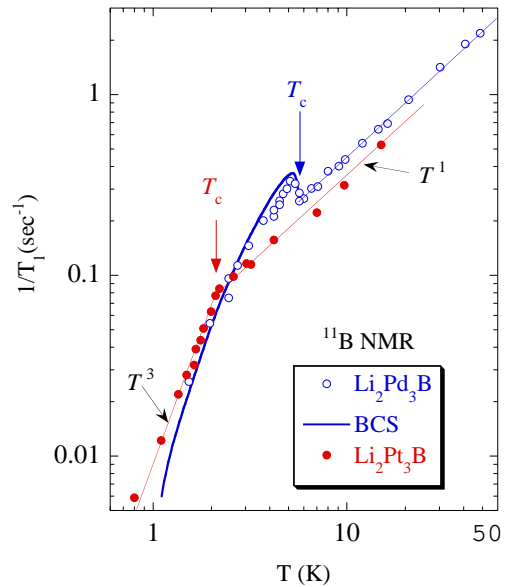}
\caption{(Color online) Temperature dependence of the $^{11}$B spin-lattice relaxation rate, $1/T_1$, in Li$_2$Pd$_3$B and Li$_2$Pt$_3$B. The experimental errors are within the circles. The arrows indicate the superconducting transition temperature $T_c$ under magnetic fields. The curve is a fit to the BCS theory with   $\Delta_0=1.1k_BT_c$. The straight lines indicate the $1/T_1 \propto  T$ and $1/T_1 \propto  T^3$ relations, respectively.}
\label{fig:3}
\end{center}
\end{figure}

In contrast, $1/T_1$ shows no coherence peak below $T_c$ in Li$_2$Pt$_3$B, and decreases as $T^3$ with decreasing temperature. This behavior indicates the existence of line nodes in the gap function, as has been seen in many    heavy fermion superconductors \cite{Maclaughlin,Kitaoka,Zheng}. The $1/T_{1S}$ in the superconducting state is expressed as
\begin{eqnarray}
\lefteqn{\frac{T_{1N}}{T_{1S}} } \nonumber \\
& = & \frac{2}{k_BT}\int\int (1+\frac{\Delta^2}{EE'})N_s(E)N_s(E') \nonumber \\
& & \times f(E)[1-f(E')]\delta(E-E')dEdE'
\end{eqnarray}
where $1/T_{1N}$ is the relaxation rate in the normal state, $N_s(E)$ is the superconducting density of states (DOS),  $f(E)$ is the Fermi distribution function and $C =1+\frac{\Delta^2}{EE'}$   is  the "coherence factor". When  the gap function has nodes, its average over the Fermi surface is zero and one obtains $C$=1. Also, the DOS at $E = \Delta$ is less divergent in such case. These two effects eliminate the coherence peak. Moreover, if the nodes form a line, the DOS at low energy is linear in $E$, which yields a $T^3$ dependence of $1/T_1$ according to eq.(1). Therefore, the $T_1$ results show that the OWF of the Cooper pair has also changed drastically from Li$_2$Pd$_3$B to Li$_2$Pt$_3$B.  Our conclusions are further supported by the $^{195}$Pt NMR results. In Fig. 4 are shown the $T$ dependence of $1/T_1$ and the Knight shift of $^{195}$Pt under a magnetic field of $H$=0.396 T. $1/T_1$ shows a $T^3$ variation below $T_c$, as in the $^{11}$B nuclear site. The Knight shift has a much larger value as expected for a transition metal element, and is invariant across $T_c$. The recent measurement of magnetic penetration depth that shows $T$-linear variation in Li$_2$Pt$_3$B is consistent with our results \cite{Yuan}. 

\begin{figure}
\begin{center}
\includegraphics[scale=0.48]{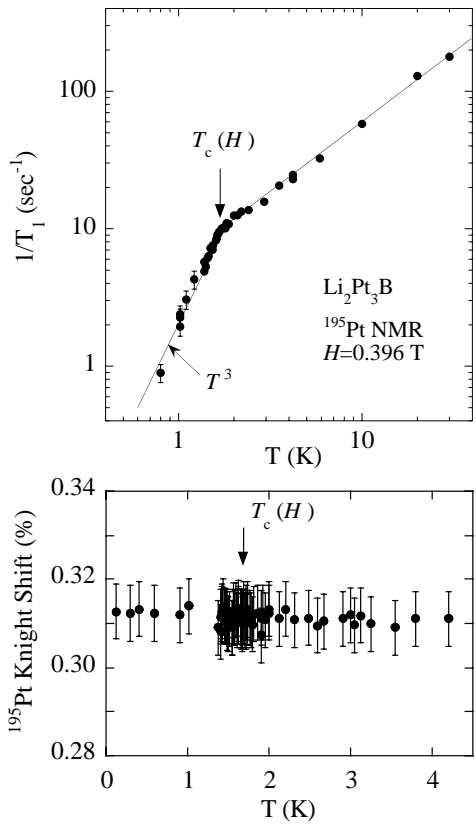}
\caption{The $T$ dependence of $1/T_1$ and the Knight shift for $^{195}$Pt in Li$_2$Pt$_3$B.}
\label{fig:4}
\end{center}
\end{figure}

A spin-triplet state with anisotropic OWF  is realized in superfluid $^3$He  \cite{Lee} and also believed to occur in  superconductors UPt$_3$  \cite{Tou}, Sr$_2$RuO$_4$  \cite{Ishida} and (TMTSF)$_2$PF$_6$  \cite{Brown}. In these cases, the existence of nodes in the gap function is a manifestation of strong electron correlations (Coulomb repulsion). However, the unconventional nature of the superconducting state of Li$_2$Pt$_3$B cannot be attributed to electron correlations. In fact, the electron correlation is as weak as in Li$_2$Pd$_3$B, as can be judged from the conservation of Korringa law, namely, $1/T_1T$ and $K_s$ are $T$-independent in the normal state (see Fig. 2 and Fig. 3). Instead, the striking difference in the pairing symmetry between Li$_2$Pd$_3$B and Li$_2$Pt$_3$B should be attributed to the difference in the SOC strength. 

The SOC is described by the Hamiltonian,
$H_{SO}= \frac{\hbar^2}{4m^2c^2}[\vec{\nabla}V(r)\times\vec{k}]\vec{\sigma}$
, where $\vec{k}$ and $\vec{\sigma}$ are the electron momentum and Pauli spin operator, respectively, and $\vec{\nabla}V (r)$ is the electrical field. The broken inversion symmetry  increases  $\vec{\nabla}V(r)$.  The magnitude of the SOC depends on the number of positive charges ($Z$) that comprise the nucleus.
As a good approximation, it goes in proportion to $Z^2$, which is about 3 times larger for Pt than Pd. 
The SOC lifts the two-fold spin degeneracy of the electron bands, violating the parity conservation. As a result, the spin-singlet and spin-triplet states are mixed \cite{Gorkov,Frigeri,Sa,Fr}. 
The extent to which the triplet-state component is mixed depends on the strength of SOC \cite{Gorkov,Frigeri,Sa,Fr}.  Our results show that Li$_2$Pt$_3$B is an extreme case in which the strong SOC makes the triplet state dominant. They explain some puzzles such as the lower $T_c$ in Li$_2$Pt$_3$B in spite of higher DOS than Li$_2$Pd$_3$B (Ref. \cite{Pickett}); generally a spin triplet state results in a lower $T_c$ for a system with otherwise same parameters \cite{Monthoux}.

 Finally, we note that a determination of the node positions in the gap function is an important issue to be settled in future works. A $p$-wave model $\Delta (\theta)=\Delta_0 cos(\theta)$, with horizontal line-nodes, will fit perfectly our $1/T_1$ data: a fitting parameter  of $\Delta_0\sim 2.5k_BT_c$ gives a  $T^3$-dependent  $1/T_1$ in the whole temperature range below $T_c$ for such model \cite{Zheng}.  On the other hand,
the  model proposed by Yuan {\it et al} \cite{Yuan} may also be able to explain the $T^3$ variation of $1/T_1$, but seems difficult to account for the lack of the coherence peak found here, since the momentum ($k$)-dependent term $\frac{\Delta(k)\Delta(k')}{E(k)E(k')}$ in the coherence factor will not cancel in such model where $\Delta(k)$ does not change sign over the Fermi surface \cite{Hayashi}. We hope that our results will stimulate more theoretical works in this regard. 


In conclusion, through extensive NMR measurements, we have found that in non-centrosymmetric Li$_2$Pt$_3$B the Cooper pair is in the spin-triplet state and there exist line nodes in the gap function. The realization of such exotic superconducting state is surprising, given that there are no electron correlations in the material. The striking difference from the isostructural Li$_2$Pd$_3$B, which is a conventional superconductor, arises from the much larger spin-orbit coupling in Li$_2$Pt$_3$B. 
Our finding shows that Li$_2$Pt$_3$B is a spin-triplet superconductor with the highest $T_c$ to date which will provide better opportunities for studying novel superconductivity. We emphasize that non-centrosymmetric superconductors are still rare, and Li$_2$(Pd,Pt)$_3$B is an ideal prototypical system for studying the effects of crystal inversion-symmetry breaking.
 

The authors wish to thank S. Fujimoto, N. Hayashi and S. C. Zhang for helpful discussions, and M. Sata for assistance in sample growing which was performed by using facilities in ISSP, University of Tokyo. This work was partly supported by a MEXT research grant  (No. 17072005).   




\begin{thebibliography}{}










\bibitem{Tsuei}
C. C. Tsuei and J. R. Kirtley, Rev. Mod. Phys. {\bf 72}, 969 (2000).

\bibitem{Co}
G. - q. Zheng {\it et al}, Phys. Rev. {\bf B 73}, 180503 (R) (2006); G. - q. Zheng {\it et al}, J. Phys.: Condens. Matter {\bf 18}, L63 (2006); T. Fujimoto {\it et al}, Phys. Rev. Lett. {\bf 92}, 047004 (2004).

\bibitem{HF}
R. H. Heffner and M. R. Norman, Comments Condens. Matter Phys., {\bf 17}, 361 (1996).

\bibitem{Lee}
D. M. Lee,   Rev. Mod. Phys. {\bf 69}, 645 (1997).

\bibitem{Tou}
H. Tou   {\it et al.},  Phys. Rev. Lett. {\bf 77}, 1374  (1996).

\bibitem{Ishida}
A. P. Mackenzie and Y. Maeno, Rev. Mod. Phys. {\bf 75}, 657 (2003).

\bibitem{Brown}
I. J. Lee {\it et al}, Phys. Rev. Lett. {\bf 88}, 017004  (2002).

\bibitem{Gorkov}
L.P.	Gorkov, and  E.I. Rashba,   Phys. Rev. Lett. {\bf 87}, 037004  (2001).

\bibitem{Frigeri}
P.A. Frigeri {\it et al.}, Phys. Rev. Lett. {\bf 92}, 097001 (2004).

\bibitem{Sa}
K.V. Samokhin, Phys. Rev. Lett. {\bf 94}, 027004 (2005).

\bibitem{Fr}
P.A. Frigeri,  D.F. Agterberg,and M. Sigrist,   New J. Phys. {\bf 6}, 115 (2004).

\bibitem{Togano1}
K. Togano  {\it et al.}, Phys. Rev. Lett. {\bf 93}, 247004 (2004).

\bibitem{Togano2}
T. Badica,  {\it et al.}, J. Phys. Soc. Jpn. {\bf 74}, 1014 (2005).

\bibitem{Bauer}
E. Bauer, {\it et al.},  Phys. Rev. Lett. {\bf 92}, 027003 (2004).

\bibitem{Akazawa}
T. Akazawa,  {\it et al.},  J. Phys.: Cond. Matter  {\bf 16}, L29 (2004).

\bibitem{Yogi}
M. Yogi {\it et al}, Phys. Rev. Lett. {\bf 93}, 027003 (2004).

\bibitem{Fujimoto}
S. Fujimoto, J. Phys. Soc. Jpn. {\bf 75}, 083704 (2006).


\bibitem{Onuki}
T. Takeuchi, Y. Onuki {\it et al}, J. Phys. Soc. Jpn. (in press).

\bibitem{Nishiyama}
M. Nishiyama,   Y. Inada, G. - q. Zheng,  Phys. Rev. B {\bf 71}, 220505(R) (2005).

\bibitem{Yokoya}
T. Yokoya {\it et al.}, Phys. Rev. B {\bf 71}, 092507  (2005).

\bibitem{Yuan}
H.Q. Yuan {\it et al.}, Phys. Rev. Lett.  {\bf 97}, 017006 (2006).
 
\bibitem{Anderson}
P. W. Anderson,  Phys. Rev. Lett. {\bf 3}, 325 (1959).

\bibitem{Slichter}
L. C. Hebel,   and C.P. Slichter,  Phys. Rev. {\bf 113}, 1504 (1959).





\bibitem{Maclaughlin}
D.E. Maclaughlin, {\it et al},  Phys. Rev. Lett. {\bf 53}, 1833 (1984).

\bibitem{Kitaoka}
Y. Kitaoka   {\it et al.},  J. Magn. Magn. Mater. {\bf 52}, 341 (1985).

\bibitem{Zheng}
G. - q. Zheng {\it et al.},  Phys. Rev. Lett. {\bf 86}, 4664  (2001).



\bibitem{Pickett}
K.-W. Lee and W. E. Pickett,  Phys. Rev. B {\bf 72}, 174505 (2005).

\bibitem{Monthoux}	
P.	Monthoux   and G. G. Lonzarich,  Phys. Rev. B {\bf 63}, 054529 (2001).

\bibitem{Hayashi}
N. Hayashi {\it et al}, Phys. Rev. {\bf B 73}, 092508   (2006).

\end{thebibliography}
\end{document}